\documentclass[sigconf,9pt]{acmart}
\usepackage{cite}
\usepackage{amsmath,amsfonts}
\usepackage[ruled, lined, linesnumbered, commentsnumbered, longend]{algorithm2e}
\usepackage{graphicx}
\usepackage{textcomp}
\usepackage{listings}
\usepackage{adjustbox}
\usepackage{multirow}
\usepackage{multicol}
\usepackage{balance}
\usepackage{subcaption}
\usepackage{makecell}
\usepackage{enumerate}
\usepackage{xcolor}
\def\BibTeX{{\rm B\kern-.05em{\sc i\kern-.025em b}\kern-.08em
    T\kern-.1667em\lower.7ex\hbox{E}\kern-.125emX}}
\usepackage[font=scriptsize]{caption}
\usepackage{array}
\usepackage{float}
\usepackage{tabularx}
\newcommand{\proj}
{\textit{UnifyFL}\xspace}

\newcommand{\diff}[1]{\noindent\textcolor{black}{#1}}

\newcommand{\reb}[1]{\noindent\textcolor{black}{#1}}

\begin{document}

\acmYear{2025}\copyrightyear{2025}
\setcopyright{rightsretained}
\acmConference[Middleware '25]{26th ACM Middleware Conference}{December 15--19, 2025}{Nashville, TN, USA}
\acmBooktitle{26th ACM Middleware Conference (Middleware '25), December 15--19, 2025, Nashville, TN, USA}
\acmDOI{10.1145/3721462.3730955}
\acmISBN{979-8-4007-1554-9/25/12}

\title{\proj: Enabling Decentralized Cross-Silo Federated Learning}


\author{Sarang S}
\affiliation{
  \institution{BITS Pilani, KK Birla Goa Campus}
  \country{India}
}
\email{f20210966@goa.bits-pilani.ac.in}

\author{Druva Dhakshinamoorthy}
\affiliation{
  \institution{BITS Pilani, KK Birla Goa Campus}
  \country{India}
}
\email{f20220131@goa.bits-pilani.ac.in}

\author{Aditya Shiva Sharma}
\affiliation{
  \institution{BITS Pilani, KK Birla Goa Campus}
  \country{India}
}
\email{f20221159@goa.bits-pilani.ac.in}

\author{Yuvraj Singh Bhadauria}
\affiliation{
  \institution{BITS Pilani, KK Birla Goa Campus}
  \country{India}
}
\email{f20201685@goa.bits-pilani.ac.in}

\author{Siddharth Chaitra Vivek}
\affiliation{
  \institution{BITS Pilani, KK Birla Goa Campus}
  \country{India}
}
\email{f20220569@goa.bits-pilani.ac.in}

\author{Arihant Bansal}
\affiliation{
  \institution{BITS Pilani, KK Birla Goa Campus}
  \country{India}
}
\email{f20200567@goa.bits-pilani.ac.in}

\author{Arnab K. Paul}
\affiliation{
  \institution{Department of CSIS and APPCAIR, BITS Pilani, KK Birla Goa Campus}
  \country{India}
}
\email{arnabp@goa.bits-pilani.ac.in}

\renewcommand{\shortauthors}{Sarang et al.}

\label{sec:abstract}
\begin{abstract}

Federated Learning (FL) is a decentralized machine learning (ML) paradigm in which models are trained on private data across several devices called clients and combined at a single node called an aggregator rather than aggregating the data itself. Many organizations employ FL to have better privacy-aware ML-driven decision-making capabilities. However, organizations often operate independently rather than collaborate to enhance their FL capabilities due to the lack of an effective mechanism for collaboration. The challenge lies in balancing trust and resource efficiency. One approach relies on trusting a third-party aggregator to consolidate models from all organizations (multilevel FL), but this requires trusting an entity that may be biased or unreliable. Alternatively, organizations can bypass a third party by sharing their local models directly, which requires significant computational resources for validation. Both approaches reflect a fundamental trade-off between trust and resource constraints, with neither offering an ideal solution. In this work, we develop a trust-based cross-silo FL framework called \proj, which uses decentralized orchestration and distributed storage. \proj provides flexibility to the participating organizations and presents synchronous and asynchronous modes to handle stragglers. Our evaluation on a diverse testbed shows that \proj achieves a performance comparable to the ideal multilevel centralized FL while allowing trust and optimal use of resources.

\end{abstract}


  
\maketitle

\section{Introduction}
\label{sec:intro}

The proliferation of Artificial Intelligence (AI) has ushered in transformative changes across diverse industries, ranging from healthcare to finance and beyond. However, the increasing integration of AI into our daily lives has given rise to a pressing concern: the potential compromise of privacy due to the extensive collection of personal data required for training these sophisticated systems. Legislations such as the General Data Protection Regulation (GDPR)~\citep{GDPR}, the California Consumer Privacy Act (CCPA)~\citep{CCPA}, the Personal Data Protection Bill (PDPB)~\citep{PDPB}, and the Personal Data Protection Act (PDPA)~\citep{PDPA} are enacted to protect the privacy of individual data. Data processors that violate these regulations are heavily penalized by regulatory agencies.

\begin{figure}[h!]
\centering
\includegraphics[width=0.7\columnwidth]{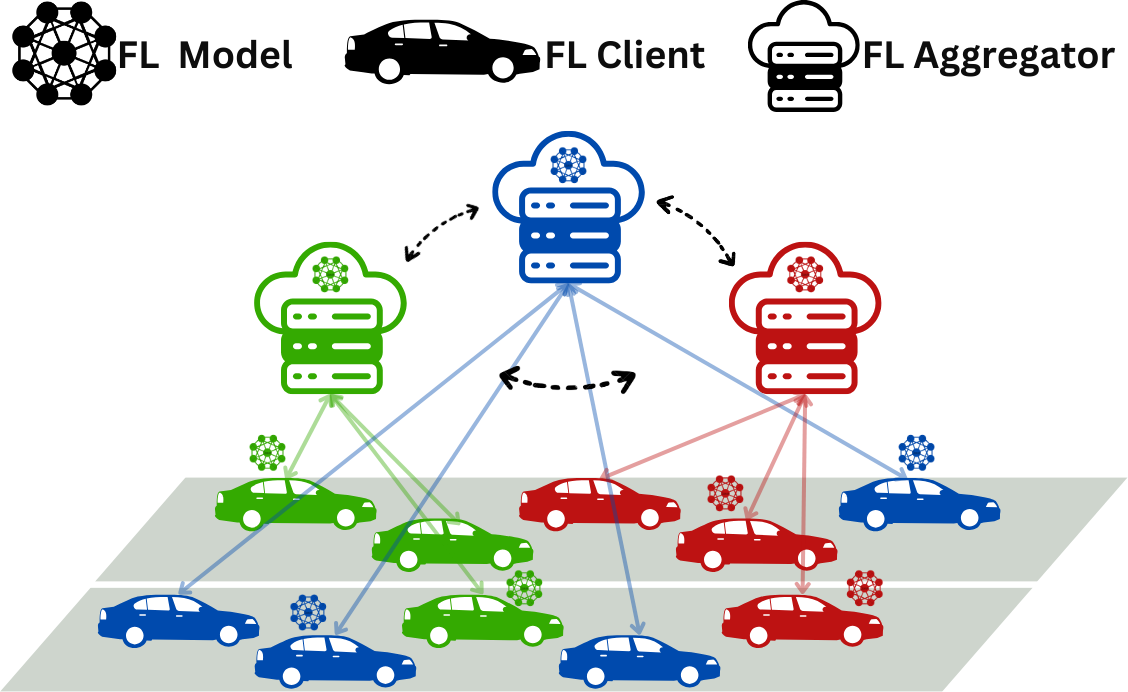}
\caption{\diff{ An illustration of a collaborative FL architecture in the automobile industry, where vehicle fleets (FL clients) from different companies train models locally and share updates with their respective FL aggregators. } }
\label{fig:vehicle}
\end{figure}

Federated Learning (FL)~\citep{mcmahan2017communication} has emerged as a promising method for addressing privacy concerns. It allows machine learning (ML) models to be trained collaboratively on multiple devices while keeping sensitive information on the user's device. Clients train their models locally and independently, sending model updates to a central server or aggregator. The aggregator merges these updates, refines the model, and sends the improved version back to the clients, completing one round of FL. The goal is to iteratively train a global model that converges to the desired accuracy. FL is applied in various areas, such as Google's GBoard~\citep{xu2023federated}, where it was first used, as well as in edge computing, recommendation systems, natural language processing (NLP), Internet of Things (IoT), healthcare, autonomous industry, and finance~\citep{shaheen2022applications}.
\subsection{Motivation}
\label{sec:Motivation}

In today's landscape, companies operating in segmented user markets often develop their own FL pipelines to enhance their models~\citep{intel_case_study}\citep{google_blog_federated_learning}\citep{owkin_federated_learning}\citep{FARAHANI2023436}\citep{zhang}. Collaborative learning among these entities can lead to substantial benefits for all the involved parties. To understand the motivation behind \proj, consider the automobile industry as illustrated in Figure~\ref{fig:vehicle}. Companies operate vehicle fleets that use sensors to collect sensitive data and use FL to train models while keeping data private on the vehicles, representing a single-level FL. However, these companies ought to collaborate to improve their models, leading to multilevel FL. Therefore, a collaborative FL architecture is essential, but must be implemented carefully to avoid disrupting existing FL systems. Two existing solutions for this are centralized aggregation and peer-to-peer aggregation.

\begin{figure}[h!]
    \begin{subfigure}{0.235\textwidth}
    \centering
        \includegraphics[width=0.7 \textwidth]{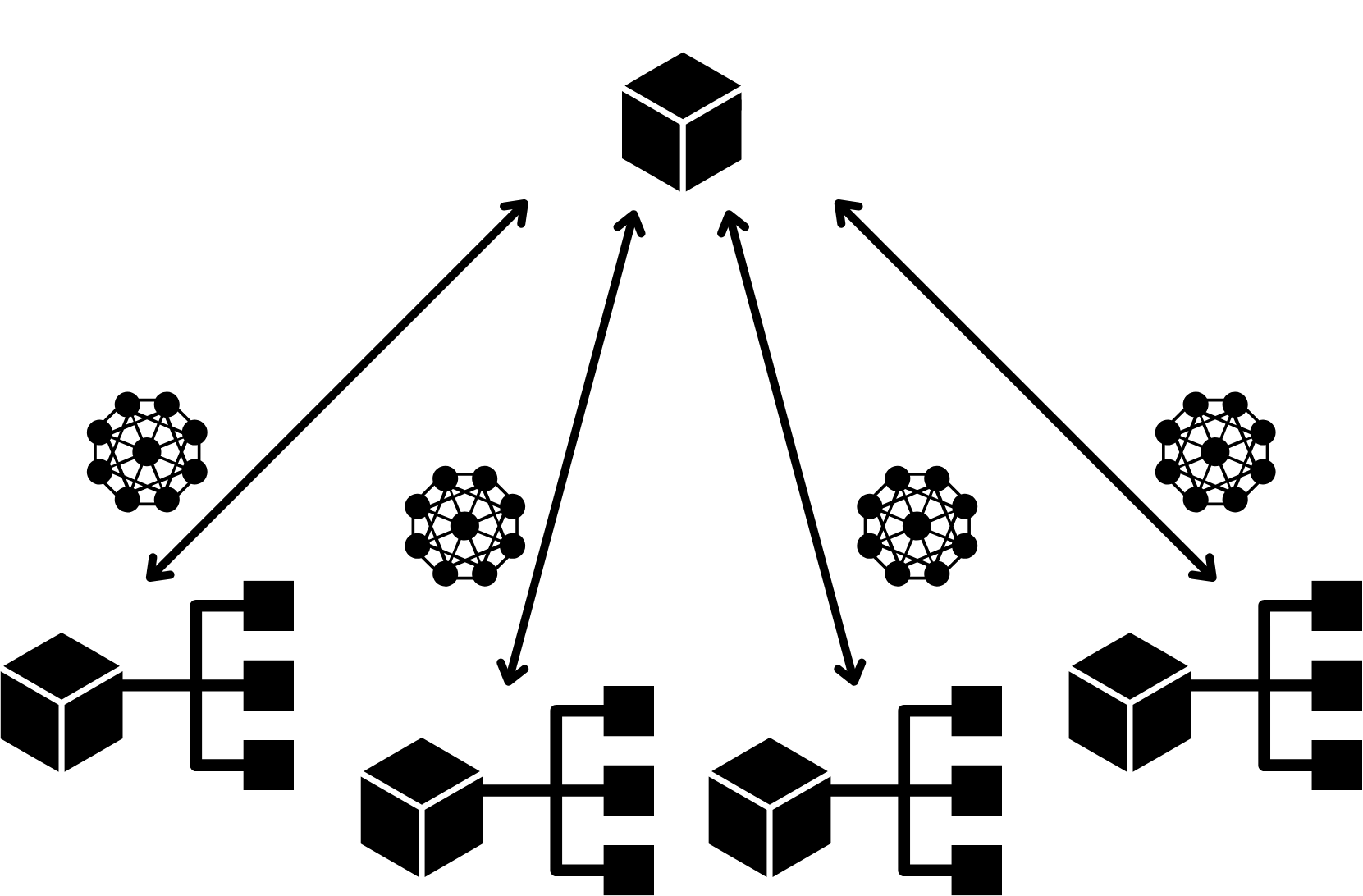}
        \subcaption{Centralized or Multilevel aggregation.}
        \label{fig:centralized}
    \end{subfigure}
    \begin{subfigure}{0.235\textwidth}
      \centering
        \includegraphics[width=0.6 \textwidth]{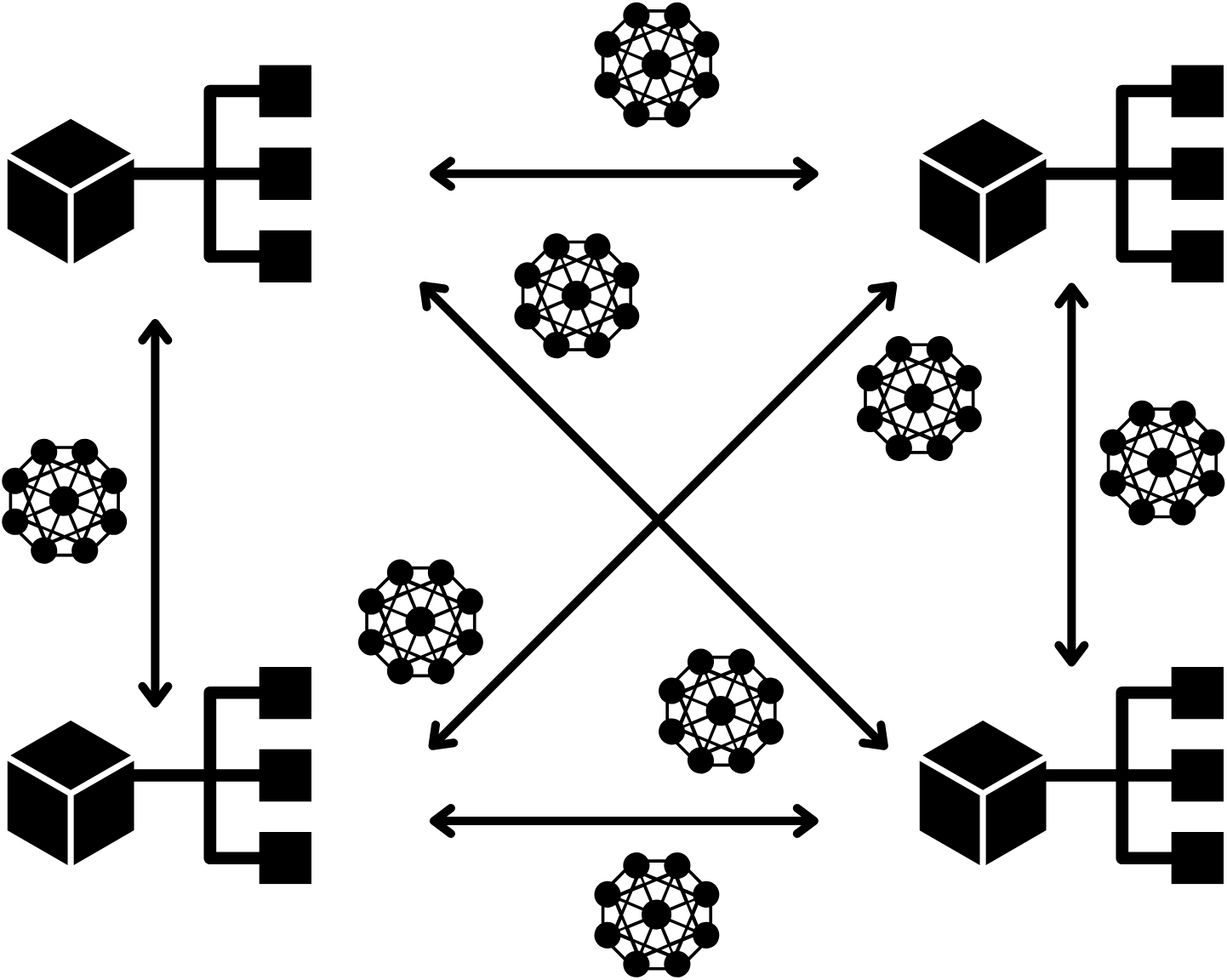}
        \subcaption{Peer-to-Peer aggregation.}
        \label{fig:p2p}
  \end{subfigure}
  \caption{Approaches for collaborative FL.}
\end{figure}

\subsubsection{A Case for Collaborative Learning}

To motivate the case for collaborative learning, we conduct two experiments of the NIID-partitioned CIFAR-10 workload on the edge cluster described in Section \ref{sec:exp_setup}. In the first experiment, the three clusters are trained independently on their respective datasets without collaboration, as in traditional FL. The results of this non-collaborative training are presented in Table \ref{tab:collab-nocollab}. However, this non-collaborative nature limits maximum accuracy to between 32.5\% - 35\% and causes overfitting due to limited and heterogeneous data. In contrast, the performance of a centralized multilevel FL approach, where clusters collaborate by sharing model updates with a central server, achieves significantly higher global accuracy of 50\% and lower global loss. This difference highlights the importance of collaboration among FL clusters, emphasizing the need for techniques that enable collaboration while preserving data privacy and addressing communication constraints.

\begin{table}
\centering
\caption{Accuracy and Loss results for No Collab and Collab settings.}
\begin{adjustbox}{width=0.3\textwidth}
\begin{tiny}
\begin{tabular}{llcc}
\toprule
 & \textbf{Cluster} & \textbf{Accuracy (\%)} & \textbf{Loss} \\
\midrule
\multicolumn{4}{l}{No Collab} \\
 & Aggregator 1 & 32.80 & 6.54 \\
 & Aggregator 2 & 35.22 & 5.97 \\
 & Aggregator 3 & 31.44 & 6.09 \\
\midrule
\multicolumn{4}{l}{Collab} \\
 & Aggregator 1 & 31.45 & 4.10 \\
 & Aggregator 2 & 34.52 & 4.73 \\
 & Aggregator 3 & 31.00 & 4.71 \\
 & Global Model & 50.4 & 1.36 \\
\bottomrule
\end{tabular}
\end{tiny}
\end{adjustbox}
\label{tab:collab-nocollab}
\end{table}

\subsubsection{Centralized Aggregation}
\label{sec:central_FL}
One approach to facilitate collaborative learning is a multilevel FL architecture with a centralized aggregator responsible for consolidating models from individual FL clusters, as shown in Figure~\ref{fig:centralized}. Centralized aggregation allows the aggregated model to learn from the entirety of the data and reduces the computational overhead for model validation. However, the reliance on a central aggregator introduces a significant trust concern and raises the risk of malicious behaviour. Each organization will have its own aggregator, but they will not reach a consensus on who holds the centralized aggregator. Moreover, participating organizations may be limited in customizing their aggregation policies and algorithms, severely restricting their ability to serve their customers better. Additionally, a malicious central aggregator could send faulty models to individual aggregators, though only targeted aggregators would be affected, while the majority would remain intact.

\subsubsection{Peer-to-Peer Aggregation}

Another approach is a fully peer-to-peer architecture, as shown in Figure~\ref{fig:p2p}, where each cluster (here, the term `cluster' is synonymous with the aggregator in the cluster) communicates directly with every other cluster, exchanging and validating models. An advantage of this approach is that it grants autonomy to the clusters to determine their aggregation strategy. However, the peer-to-peer model introduces network overhead due to the constant transfer of models. Moreover, the organizations do not have a set of rules they must adhere to, leading to easier manipulation by malicious organizations. In particular, malicious aggregators may send faulty models to honest aggregators, thus increasing computational overhead for model verification.

\begin{figure}[h!]
    \centering
\includegraphics[width=0.8\columnwidth]{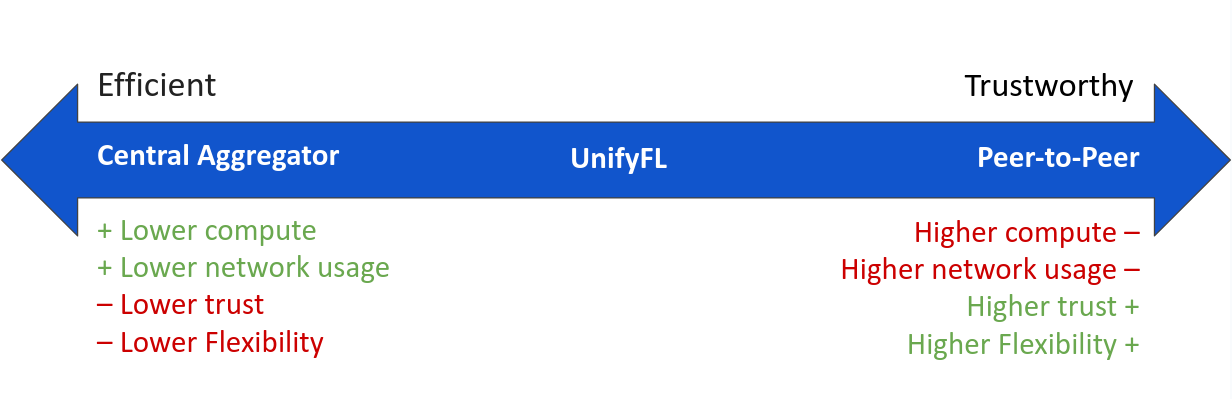}
   \caption{Centralized (or Multilevel) aggregation vs Peer-to-Peer aggregation.}
  \label{fig:trade-off}
\end{figure}

\subsubsection{The Building Blocks}

A trade-off exists between efficiency and trustworthiness when considering centralized versus peer-to-peer aggregation models, as depicted in Figure \ref{fig:trade-off}. The centralized approach leans towards efficiency, while the peer-to-peer approach emphasizes trustworthiness. Importantly, the notion of "trustworthiness" in Figure~\ref{fig:trade-off} assumes that a majority of clusters are honest, allowing independent aggregation of a reliable model. This contrasts with centralized systems, where the aggregator holds full control, introducing a single point of failure. Striking a balance between these two extremes is crucial for creating a collaborative FL architecture. To construct such a framework, two essential building blocks are required, namely a decentralized orchestrator and a distributed storage system, to support scalability, security, and seamless collaboration across multiple clusters. The decentralized orchestrator manages model training and validation, and coordinates storage and access to the latest model updates. The distributed storage ensures accessibility and reliability of model weights while ensuring immutability. Together, they form the backbone of a robust framework that reconciles efficiency with trustworthiness, enabling decentralized cross-silo FL at scale.

\subsubsection{\proj: Bridging the Gap}
Our proposed architecture, \proj, provides a balanced solution by leveraging a decentralized orchestrator and distributed storage to optimize privacy, security, and resource efficiency. Designed to enhance the privacy and security features of FL, \proj supports flexible aggregation strategies while minimizing computational overhead. By effectively balancing efficiency and trustworthiness, it enables seamless collaboration among diverse FL clusters without compromising the integrity of individual organizations' FL pipelines. Specifically, our implementation is built on the Flower FL framework~\citep{beutel2020flower}, a private Ethereum~\citep{Buterin2013} blockchain-based orchestrator and a local IPFS~\citep{ipfs} for distributed storage. \proj leverages blockchain and IPFS to ensure all aggregators receive identical models, enhancing transparency. Additionally, randomized scorers publicly share their scores, reducing scoring burdens on the aggregators while allowing model aggregation decisions to remain fair and auditable. An in-depth reasoning for these choices is provided in Section~\ref{sec:impl}.

\subsection{Key Contributions}  
This work introduces a novel framework, \proj, which bridges the gap between efficiency and trustworthiness in decentralized cross-silo FL. Our key contributions are:  
\begin{itemize}  
\item Design a cross-silo FL framework that fosters trust and collaboration among diverse FL clusters through the integration of a decentralized orchestrator and distributed storage.  
\item Implement the proposed framework using Ethereum for the decentralized orchestrator, IPFS for distributed storage and integrate it with the widely used Flower FL framework to allow seamless adoption in existing FL pipelines while ensuring scalability and security.
\item Introduce two modes of operation - \textit{synchronous} mode, where all aggregators operate in sync every round, and the \textit{asynchronous} mode, which accommodates stragglers to improve resource utilization and idle time.
\item Develop a scoring mechanism to evaluate models received from aggregators using various algorithms, removing bias and enabling trust in model aggregation.  
\item Enable high customizability, allowing aggregators to adopt different aggregation policies, scoring mechanisms, and training configurations, offering unparalleled flexibility in FL.  
\item Evaluate \proj on real-world testbeds under diverse configurations to demonstrate its effectiveness in achieving high accuracy, efficient resource utilization, and robust trust mechanisms.  
\end{itemize}  

The subsequent sections of this paper are organized as follows: Section \ref{sec:background} provides background information on types of Federated Learning (FL), Blockchain, IPFS and other relevant topics. In Section \ref{sec:system-design}, we delve into the system design of \proj, exploring its individual components, orchestration phases, modes of operations and implementation. Section \ref{sec:eval} conducts a thorough analysis of the performance of \proj{}, Section \ref{sec:discussion} discusses current challenges and potential future directions, and Section \ref{sec:conclusion} concludes the paper.

\section{Background \& Related Work}
\label{sec:background}

Cross-Device FL applies to scenarios involving a large number of edge devices in a single organization. Cross-Silo refers to FL among a number of organizations. 

\subsection{Cross-Device FL}

Bao \textit{et al.} introduce \textit{FEDCOLLAB}~\citep{bao2023optimizing}, an FL framework that mitigates negative transfer by clustering clients into non-overlapping coalitions based on data distribution distances and quantities. This ensures collaboration primarily among similar clients, expanding partnerships when data is scarce. Wang \textit{et al.} propose \textit{PTDFL}~\citep{wang2023enhancing}, a decentralized FL scheme enhancing privacy and trustworthiness via gradient encryption and local aggregation, eliminating the need for a trusted third party. Wahrstätter \textit{et al.} introduce \textit{OpenFL}~\citep{wahrstatter2024openfl}, an autonomous Ethereum-based smart contract platform with a collateral-backed reputation system to mitigate attacks.

While \textit{FEDCOLLAB}, \textit{PTDFL}, and \textit{OpenFL} excel in single-level FL, their architectures limit applicability to multilevel FL. \proj{} addresses these complexities by supporting hierarchical relationships among clients and global models.

\subsection{Cross-Silo FL}

Goh \textit{et al.} propose a Blockchain-based FL (BCFL) reference architecture~\citep{goh2023blockchain} incorporating Flower, Ethereum, and IPFS. While extending to cross-silo and cross-device scenarios, it lacks multilevel FL, diverse orchestration modes, and flexible aggregation policies. Sarhan \textit{et al.} introduce $HBFL$~\citep{sarhan2022hbfl}, \reb{a hierarchical FL framework with three layers: edge (training), combiner (aggregation), and reducer (final aggregation on blockchain)}. However, $HBFL$ centralizes power in the reducer, restricting combiners’ flexibility. In contrast, \proj{} enables organizations to choose aggregation criteria while securing models via a decentralized database. Yuan \textit{et al.} present $ChainFL$~\citep{yuan2021secure}, a two-layered BCFL with subchain (synchronous) and mainchain (asynchronous) aggregation to reduce resource consumption. However, $ChainFL$ enforces strict client weight considerations and uses a shared test dataset, compromising FL’s privacy-preserving nature. \proj{} allows aggregators to freely select aggregation modes, ensuring greater flexibility.

\begin{table}[h!] \caption{\diff{Table comparing the different FL frameworks.}} \begin{adjustbox}{width=\columnwidth} \begin{tabular}{ |c|c|c|c|c|c| } \hline \textbf{Framework} & BCFL ~\citep{goh2023blockchain} & HBFL~\citep{sarhan2022hbfl} & ChainFL ~\citep{yuan2021secure} & \textbf{\proj} \\ \hline \textbf{FL} & Single-level & Hierarchical & Hierarchical & Hierarchical\\ \textbf{Type} & Cross-Device & Cross-Silo & Cross-Device & Cross-Silo \\ \textbf{Orchestration} & Sync & Sync & Sync & Sync and Async \\ \textbf{Flexibility} & None & None & None & Flexible \\ \hline \end{tabular} \end{adjustbox} \label{tab:frameworks} \end{table}

The different frameworks are summarized in Table~\ref{tab:frameworks}. \textbf{FL} denotes whether a framework is single-level (central aggregator) or hierarchical (multiple aggregator levels). \textbf{Type} differentiates Cross-Silo vs. Cross-Device FL. \textbf{Orchestration} defines whether clusters train simultaneously (sync) or upon receiving updates (async). \textbf{Flexibility} indicates whether aggregators score and aggregate models independently.

\subsection{Blockchain}

Blockchain, a decentralized ledger technology, is the backbone for various applications, offering features such as decentralization, traceability, anonymity, and immutability~\citep{nakamoto2008bitcoin}. Blockchain employs consensus mechanisms like Proof of Work (PoW) ~\citep{nakamoto2008bitcoin} and Proof of Authority (PoA)~\citep{szilágyi2017eip} to validate and agree upon immutable blocks containing transaction data. Some blockchains incorporate smart contracts, which are self-executing agreements with predefined logic, making them suitable as decentralized orchestrators.

Blockchains can be classified into two distinct types: private and public. Public blockchains are open, permissionless networks that offer high transparency and decentralization, suitable for applications that require large-scale public trust, such as decentralized finance, cryptocurrencies and more. Private blockchains are closed, permissioned networks controlled by a single / group of entities, ideal for enterprise applications and inter-organizational data sharing. It provides better performance, privacy, and control, but they rely on trust in the controlling entities. Our implementation of \proj uses a private Ethereum-based blockchain with smart contracts and uses Clique~\citep{szilágyi2017eip} PoA as the consensus protocol to reduce resource utilization.

\subsection{InterPlanetary File System (IPFS)}

The InterPlanetary File System (IPFS)~\citep{ipfs} represents a decentralized storage system that fragments data into smaller, distributed pieces across a network of connected nodes. Each file stored in IPFS gets assigned a Content Identifier (CID) when it gets fragmented and stored on the filesystem, which ensures data integrity and security while improving data availability by reducing the reliance on single storage entities. With its desirable features of content-addressing for data integrity, peer-to-peer architecture eliminating single points of failure, and ease of setup and interaction for individual nodes, IPFS provides a compelling combination of resilience, security, and user-friendly decentralization and stands out as the ideal decentralized storage for \proj. 

\subsection{IID and Non-IID Data}

In machine learning, data distributions can be classified based on their underlying assumptions.
Independent and Identically Distributed (IID) data assumes that the probabilities of all random variables are independent and mutually exclusive, ensuring that each client's data is a representative sample of the entire dataset. In contrast, non-IID (NIID) partitioning introduces a more realistic and challenging setting where data is skewed or siloed across clients. In FL and other distributed training approaches, it is crucial to have well-balanced data distributions across clusters. This ensures each cluster can learn generalized patterns effectively instead of becoming biased toward specific features in their limited data. As shown via experiments, \proj can handle both IID and NIID data.

\subsection{Scoring Algorithms}
\label{sec:scoring}

Scoring algorithms play a vital role in FL by helping the aggregator assess the quality of each client's model update. These scores are crucial for identifying and filtering out bad updates, particularly in the context of poisoning attacks, where a malicious client tries to degrade the model's performance. By evaluating model updates through a scoring system, the aggregator can minimize the impact of harmful updates, ensuring that only high-quality contributions are included in the aggregation process. To mitigate these attacks, various scoring algorithms have been developed with different properties, like using accuracy as mentioned in BlockFlow~\citep{mugunthan2020blockflow} and computing model similarity as in $MultiKRUM$ \citep{peyvandi2022privacy, shayan2020biscotti, zhao2020privacy}.  

Accuracy can be used as a metric to weed out poorly performing models and works with both asynchronous and synchronous training modes, but it requires a testing dataset on each scorer and is computationally heavy. $Multikrum$, on the other hand, is an easily computable similarity-based score but can only be used in synchronous training mode since it operates on all the models submitted in a round and scores them at once. 

\proj~ supports multiple scoring functions with different compute and time requirements. Both $MultiKRUM$ and accuracy-based scoring are added to our implementation. For evaluating \proj, accuracy-based scoring is primarily used as it supports both asynchronous and synchronous orchestration modes and provides a more accurate comparison between the two modes.

\section{System Design of \proj}
\label{sec:system-design}

\begin{figure*}[h]
\centering
\includegraphics[width=0.6\linewidth]{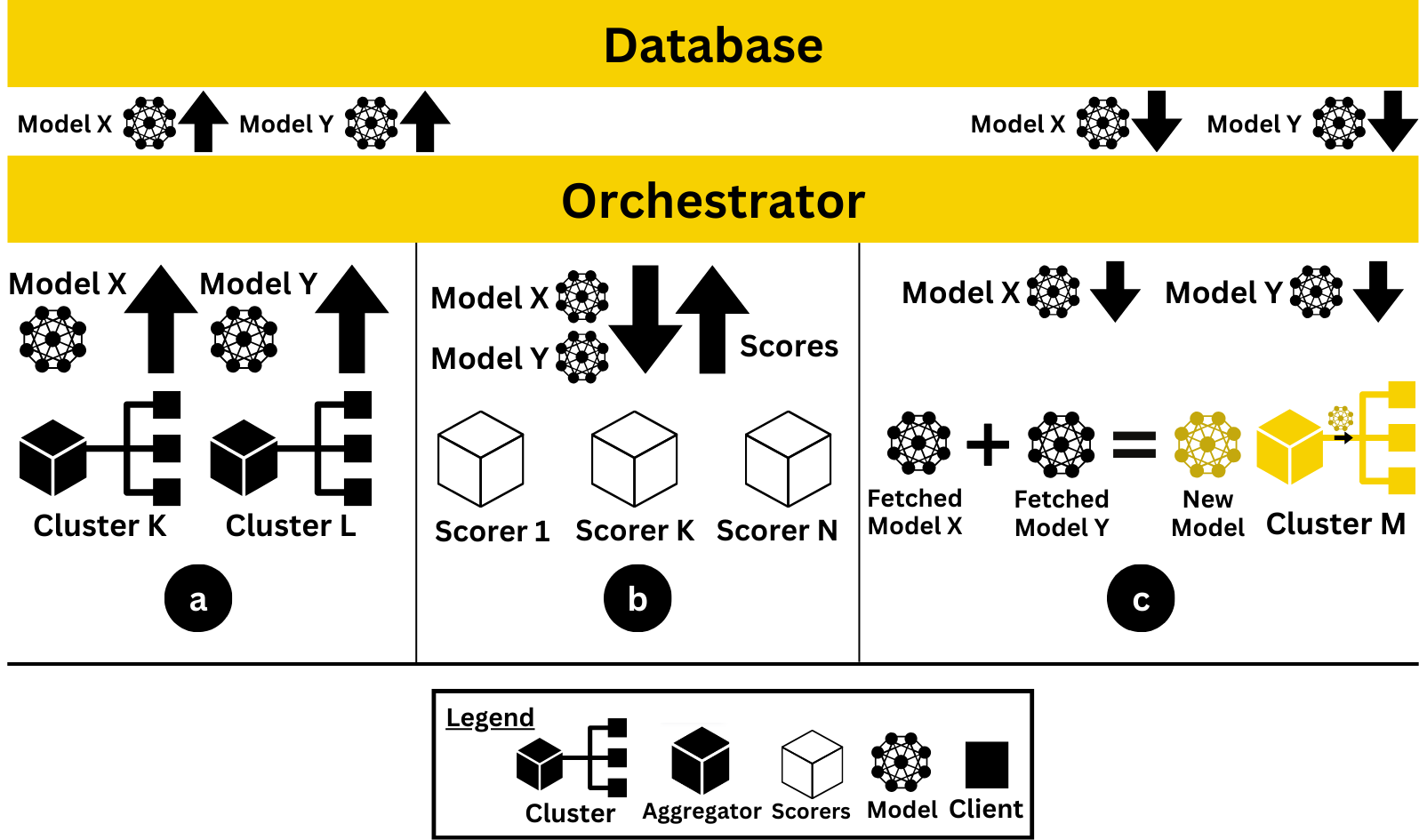}
   \caption{A step-by-step walk-through of \proj Framework. (a) Clusters perform local training, aggregate the results, and submit local weight for scoring. (b) Scorers pull weights and submit their scores.  (c) Aggregators pull and aggregate global models to send to clients for the next round. }
  \label{fig:design}
\end{figure*}

The high-level architecture of \proj{} with the step-by-step description is shown in Figure~\ref{fig:design}. 


\proj revolves around three main components:
\begin{itemize}
    \item Participating \textit{FL clusters} willing to share their aggregated models and act as \textit{scorers} for other models.
    \item An \textit{orchestrator} that coordinates the sharing of aggregated model weights and manages the scoring of submitted models to verify these weights.
    \item A \textit{distributed database} or storage that securely stores aggregated model weights, accessible to all participating FL clusters.
\end{itemize}

These components interact seamlessly through a well-defined workflow through six significant steps.

\begin{enumerate}
    \item Once the aggregation of local models received from clients of cluster $K$ is completed, the aggregated model weight $X$ is stored on the database, and the address is sent to the orchestrator. Similarly, model $Y$ from cluster $L$ as shown in Figure~\ref{fig:design} (a).
    \item Upon receiving the address of model weights $X$ and $Y$, the orchestrator selects from the list of $N$ aggregators; a majority of ($N/2 + 1$) scorers to score the model weights $X$ and $Y$ for verification. This majority is chosen to remove bias.
    \item These scorers score the model weight $X$ using the chosen scoring function. The obtained scores are sent to the orchestrator, as shown in Figure~\ref{fig:design} (b).
    \item When an aggregator, such as cluster $K$, desires to push a global model to its clients for the next round of local training, it can query the orchestrator to retrieve information about the aggregated model weights available on the database, along with their corresponding scores.
    \item Based on its criteria for model usefulness, the aggregator can pull a model and aggregate it with its current global model as shown in Figure~\ref{fig:design} (c). 
    \item The newly aggregated global model is subsequently sent to the clients for the next round of local training.
\end{enumerate}

This orchestrated workflow ensures secure, transparent, and efficient knowledge transfer among FL clusters while preserving the privacy and autonomy of individual implementations. In the following sections, we delve into the intricate details of each component and step, highlighting the design choices that make \proj{} a transparent and collaborative framework for FL.

\subsection{Orchestration Phases}

In \proj{}, the framework components interact in two major phases. We explain the two phases in more detail below. 

\subsubsection{Training Phase of Aggregators} 
When an aggregator of a cluster wants to start a round of local training, it queries the orchestrator to know the weights available on the database with their scores. The aggregators pull the aggregated model weights of the other aggregators from the distributed database. These weights are then aggregated with the aggregator's current local model and sent to the aggregator's clients for local training. When one round of local training is completed, the aggregator aggregates the weights from its clients, forming the local model, saves it to the database, and submits it to the orchestrator for scoring.

\subsubsection{Scoring Phase of Scorers} 
Once the orchestrator receives the weight, it randomly selects a majority subset of scorers from the pool of registered scorers to score the model and submit the scores. The scorers then pull the assigned models, score them using the designated scoring algorithms like accuracy and $MultiKRUM$ (Section \ref{sec:scoring}), and send the scores to the database.\\ 

\textbf{Duality of a cluster: }Since a cluster will function both as the aggregator of local weights and the scorer of other aggregator's models, \proj can undergo two modes of synchronization - synchronous mode (\textit{Sync}) - where the two phases of all clusters are in sync, and the asynchronous mode (\textit{Async}) - where clusters can start their training and scoring phases without having to wait for other clusters.

\subsection{Synchronous Mode (\textit{Sync})}


In the \textit{Sync} mode, the orchestrator notifies all the aggregators to start the training phase together and opens the window for the submission of models. Once the window closes, it sends the models to the scorers and opens a similar window for submission of scores. All the components work synchronously. The disadvantage of this mode is that there will be idle time for the fast aggregators that wait for the phase to end once they have submitted the model. There could be a straggling aggregator that cannot submit its local model weights during the present round. This could be due to the aggregator's lesser system capabilities or local training on many clients with huge datasets. Either way, the weights by the straggler are important and thus will be allowed to be submitted in the next round.

In synchronous orchestration within \proj{}, the workflow is meticulously divided into distinct phases, forming a cycle. Each crucial step is allocated to a specific phase duration, requiring an aggregator to adhere to the predefined schedule. Even if an aggregator completes its aggregation process during the pulling or scoring phase, it must patiently wait for the submission phase to submit its aggregated model officially. The orchestrator systematically cycles these phases, and clear signals about the initiation or conclusion of each phase are disseminated to all participating aggregators.

\begin{figure}[t]
\centering
\includegraphics[width=1\columnwidth]{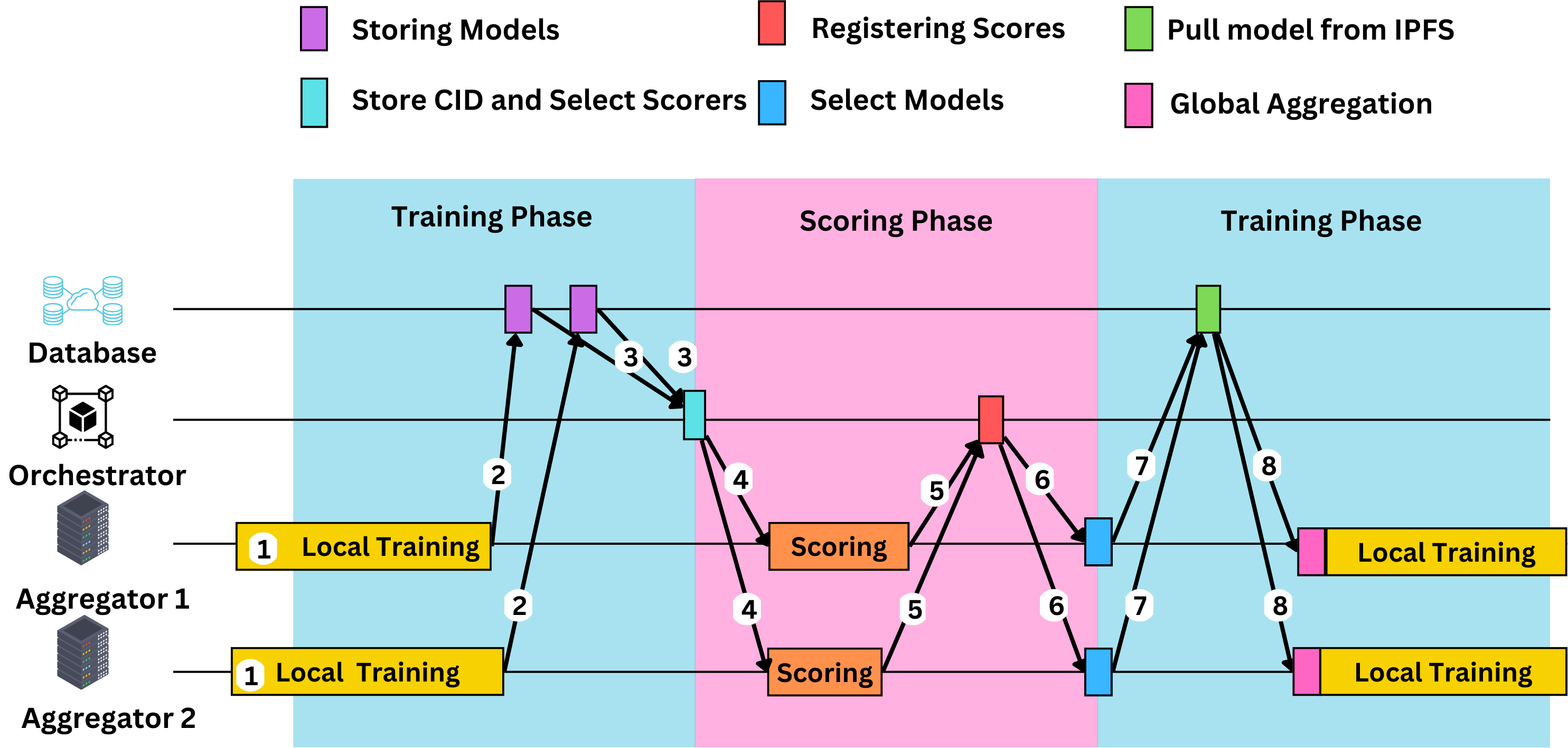}
      \caption{Workflow of the \textit{Sync} mode in \proj{} with two aggregators.}
  \label{fig:syncdesign}
\end{figure}

Figure~\ref{fig:syncdesign} illustrates the Sync Mode workflow, where the orchestrator waits for the Training Phase to conclude before initiating the Scoring Phase, assigning $n/2+1$ scorers to evaluate models received in the current round.

\textbf{Stragglers: }
As outlined above, straggling may happen when the designated local training, aggregation, or scoring tasks exceed their allocated duration in their respective phases. Should an aggregator fail to submit the aggregated model weights within the Submission Phase due to late completion of either client local training or model weight aggregation at the aggregator, it must wait for the next phase to submit. Similarly, if there is a delay in scoring, and the aggregator that acts as a scorer misses the current scoring phase, the blockchain will no longer accept scores. Straggling can cause either a delay in the submission timeline or a loss of computational resources in scoring. To counter these effects, we propose an \textit{Async} orchestration mode to tackle the challenges of straggling. This approach seeks to alleviate delay impacts by allowing clusters to function independently, thus reducing the wait time for lagging components and enhancing the overall efficiency of the FL system.

\subsection{Asynchronous Mode (\textit{Async})}
In \textit{Async} mode, every aggregator autonomously starts its training phase. This ensures minimal idle time for the aggregators. The models are also sent to the scorers as they get submitted contrary to the \textit{Sync} mode. We aim to maximize resource utilization in \proj by making the orchestrator intelligent. The orchestrator can make decisions about the availability of an aggregator based on whether the aggregator has submitted local weights or scores. In \textit{Async} \proj, the training process is not synchronized, thus allowing clients to independently perform local training and communicate with the central server whenever they have updated model parameters. This lack of synchronization allows for more flexibility in resource allocation and is particularly suitable for heterogeneous devices. 

\begin{figure}[h]
\centering
\includegraphics[width=1\columnwidth]{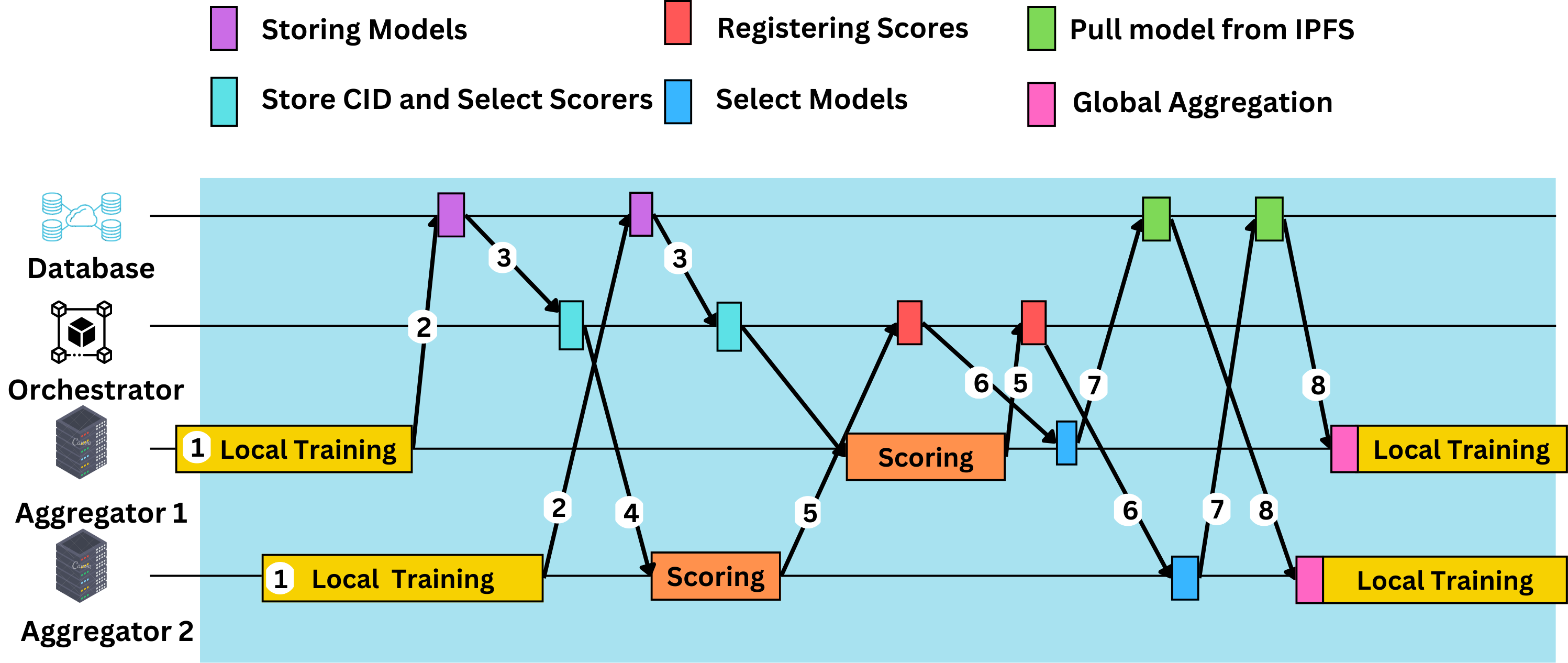}
\caption{Workflow of \textit{Async} mode in \proj{} with two aggregators.}
\label{fig:asyncdesign}
\end{figure}

\diff{ Figure~\ref{fig:asyncdesign} depicts the Async Mode workflow with staggered round starts for the two aggregators, where the orchestrator immediately assigns scorers from idle aggregators to evaluate models as soon as their CIDs are submitted. }

The example workflow of \proj is shown in Figure~\ref{fig:syncdesign} (Sync Mode) and Figure~\ref{fig:asyncdesign} (Async Mode) with two aggregators. The steps involved are as follows.

\begin{enumerate}
    \item After local training, the local weights from the clients in that cluster get aggregated by the local aggregator.
    \item Aggregated local model gets stored in IPFS.
    \item The smart contract in the blockchain-based orchestrator is updated with the Content Identifier (CID) of the model on IPFS.
    \item In \textit{Sync} mode, the orchestrator waits for the Training Phase to end. Once the Scoring Phase begins, the smart contract assigns \(n/2 + 1\) scorers to score each of the models it received in the current round. It sends the scorers the CIDs of the models received in the current round of training. However, in \textit{Async} mode, the smart contract assigns the scorers from idle aggregators as soon as the model's CID is submitted.
    \item The scorers pull model weights from IPFS, score them with their test set and submit their scores to the smart contract. The smart contract maps and stores the CID and the list of scores received from the scorers.
    \item The smart contract sends the CIDs and the scores of the models to the aggregators. In the case of \textit{Sync} mode, this is performed once the Scoring Phase ends and the orchestrator disregards any scores submitted after this phase.
    \item The aggregators may then use their policies to determine which models to use for the next round of training and request to pull the model weights from IPFS using the CID received from the smart contract.
    \item IPFS responds with the model weights and aggregates them by the aggregator before local training starts.
\end{enumerate}

\diff{ \textbf{\textit{Sync} vs \textit{Async}: }
The key differences between \textit{Sync} and \textit{Async} modes are summarised in Table \ref{tab:modes}. In the \textit{Sync} mode, training and scoring phases start simultaneously for all the clusters, with all clients required to submit their weights before proceeding, leading to high idle times and susceptibility to delays caused by stragglers. It also ensures access to weights from all clients and supports weight similarity-based scoring. In contrast, the \textit{Async}  mode operates independently for both phases, does not require waiting for all weights, and is less affected by stragglers, resulting in lower idle times. However, \textit{Async}  mode does not necessarily guarantee access to all weights, which can lead to lower accuracy and does not support weight similarity-based scoring. }

\begin{table}[h!]
  \caption{Comparison between synchronous ($Sync$) and asynchronous ($Async$) modes.}
  \label{tab:modes}
  \scriptsize
  \begin{tabular}{lll}
    \toprule
    Property & $Sync$ & $Async$\\
    \midrule
    Training phase start & Together& Independent\\
    \midrule
    Scoring phase start & Together& Independent\\
    \midrule
    \makecell[l]{Awaiting submission \\ of all weights} & Yes & No\\
    Impact due to stragglers & High& Low\\
    \midrule
    \makecell[l]{Access to weights \\ from all clients}& Necessarily & Not necessarily\\
    Idle time & High & Low\\
    \midrule
    \makecell[l]{Weight similarity-based \\ scoring} & Supported & Not supported\\
    \bottomrule
\end{tabular}
\end{table}

\subsection{Implementation of \proj}
\label{sec:impl}
Here, we explain the implementation details of \proj.

\subsubsection{Decentralized Orchestrator}

Centralized cloud solutions, such as traditional compute platforms, are often used as orchestrators to run and manage tasks across distributed systems. These systems provide high performance and scalability, making them practical for coordination. However, they rely on a single provider or central authority, introducing risks such as a single point of failure, lack of transparency, and over-reliance on external infrastructure. In contrast, a decentralized orchestrator, like the one in \proj to manage the training activities of multiple aggregators and foster seamless collaboration, eliminates these risks by distributing control, enhancing resilience, and fostering trust among participants without relying on third-party providers. The orchestrator plays a pivotal role in directing aggregators to store models on the decentralized database, collecting scores for the models from scorers, and ensuring access to the latest models by all aggregators. 

In our implementation of \proj, the decentralized orchestrator is realized through a private chain of Geth (go-ethereum)~\citep{Buterin2013} nodes, featuring smart contracts developed in Solidity. Ethereum is the best pick for this implementation due to its mature ecosystem, robust security features, and extensive developer support, which enable reliable and scalable smart contract execution on a decentralized network. The chain leverages the Clique algorithm, implementing Proof of Authority (PoA) as the consensus mechanism. PoA is chosen to provide high security, scalability with minimal computing power consumption, and faster transaction validation. 


The participating aggregators with distinct internal implementations and datasets yield aggregated models with varying accuracies, aggregation time durations, and completion times. The decentralized orchestrator addresses these diversities and ensures collaboration among all aggregators. It performs critical functions, including directing an aggregator to store the aggregated model on the decentralized database, validating aggregated models before permitting storage, informing aggregators about the availability and scores of aggregated model weights, and directing aggregators to fetch selected aggregated models from the decentralized database.

\subsubsection{Distributed Database}

A shared immutable database is essential to store the model weights generated continuously by training ML models across authorized aggregators, ensuring not only accessibility but also safeguarding against malicious modifications. Decentralized databases emerge as an optimal solution for this purpose due to their inherent characteristics of immutability and transparency. Unlike centralized databases, decentralized databases are distributed, allowing for scalability by adding more nodes. Notably, we can leverage this advantage by hosting decentralized database nodes on the aggregator nodes of aggregators, eliminating the necessity for extra nodes dedicated solely to storage.

In our implementation of \proj, we have chosen IPFS as our decentralized database solution as it provides an efficient method for sharing model updates across geographically distributed networks, making it a suitable choice over other distributed storage options like parallel file systems and cloud storage. Parallel file systems are not viable since aggregators do not need to be co-located, while cloud storage is unsuitable because it requires distributed ownership between organizations, necessitating more complex tools. IPFS stores data in smaller, distributed pieces across a network of connected nodes, employing content-based addressing to ensure data integrity and security. These features collectively contribute to the robustness and scalability of our system, aligning seamlessly with the requirements of a decentralized and secure FL environment.

\begin{algorithm}[h!]
\scriptsize
\DontPrintSemicolon
\SetKwFunction{Fun}{\textbf{startTraining}}
    \Fun{}{:}{
    
        \quad{Starts the training phase.}
        \texttt{\\}
        \quad{Emits $StartTraining$ event that aggregators subscribe to.}
    }
\texttt{\\}
\SetKwFunction{Fun}{\textbf{submitModelValidTrainer}}
    \Fun{model}{:}{
        \texttt{\\}
        \quad{Called by an aggregator to submit a model CID to the orchestrator.}
    }
\texttt{\\}
\SetKwFunction{Fun}{\textbf{startScoring}}
    \Fun{}{:}{
        \texttt{\\}
        \quad{Samples a majority subset of scorers to assign the model scoring.}
        \texttt{\\}
        \quad{Emits $StartScoring$ event that aggregators subscribe to.}
        }
\texttt{\\}
\SetKwFunction{Fun}{\textbf{submitScoreValidScorer}}
    \Fun{model, score}{:}{
        \texttt{\\}
        \quad{Called by a scorer to submit a score to the orchestrator.}
        }
\texttt{\\}
\SetKwFunction{Fun}{\textbf{getLatestModelsWithScores}}
    \Fun{}{:}{
        \texttt{\\}
        \quad{Returns the latest set of models with a set of scores for each model.}
    }

    \caption{Smart Contract in \proj.}
    \label{algorithm}
\end{algorithm}

\subsubsection{Smart Contracts}

The orchestrator is implemented in Solidity smart contracts deployed on the Ethereum blockchain. The smart contract used for the orchestration is shown in Algorithm \ref{algorithm}. 

The \texttt{startTraining()} function initiates the training phase and emits an event to notify the aggregators of the start of the training. After completing the training, the models are submitted to the orchestrator. The orchestrator records the submitted models and emits an event to inform the external parties about the model submission. The \texttt{startScoring()} function initiates the scoring phase and generates a random list of participants eligible for scoring. The valid scores are submitted to the orchestrator for specific rounds and models. The orchestrator records the scores and emits an event to inform the external aggregators about the score submission.

\subsubsection{Policies}
\label{sec:policies}
The smart contract sends all the global models, each with their respective scores from multiple scorers. In order to pick the ideal models to aggregate and use for training, two sets of policies are required: one for picking the list of models to be aggregated and one for corroborating a score with each model.

\par{\textbf{Aggregation Policies: }}In \proj{}, various model aggregation policies are implemented to govern the selection and combination of aggregated model weights from different FL clusters. The selected policy determines how scores associated with these models are considered during aggregation. The following policies are implemented for the evaluation:

\begin{itemize}
    \item \textbf{Sampling-based Policies:} These policies involve various methods for selecting models for aggregation. 
    \begin{itemize}
        \item \textit{Random k} (randomly selecting k models)
        \item \textit{All} (aggregating all models)
        \item \textit{Self} (considering only the local model)
    \end{itemize}
    
    \item \textbf{Performance-based Policies:} These policies focus on selecting the top-performing models based on scores:
    \begin{itemize}
        \item \textit{Top k} (selecting the top k models)
        \item \textit{Above Average}
        \item \textit{Above Median}
        \item \textit{Above Self}
    \end{itemize}
\end{itemize}

\proj leaves the choice of the number models to be selected ($k$ in $Random$ $k$ and $Top$ $k$) to the aggregators.

\par{\textbf{Scoring Policies: }}All the performance-based policies rely on the model scores to determine which models to pick for aggregation. However, the smart contract provides a list of scores for each model, each provided by different scorers. Several scoring policies are implemented for each use case to use the global models effectively. Some policies include selecting the median, minimum, and maximum scores. In the case of scorers that do not perform as expected due to malicious intents or the split of the scorers' testing datasets, these policies will play an important role in determining the model chosen for aggregating. These policies also provide flexibility and adaptability to cater to diverse FL scenarios, allowing organizations to tailor the aggregation strategy according to their specific requirements and priorities. The ability to seamlessly switch between policies or implement a custom approach enhances the versatility of \proj{} in accommodating a wide range of use cases.

\subsubsection{End-to-End implementation}

We implement the end-to-end \proj fframework on top of the widely used Flower FL framework. The server implementation is extended to incorporate calls to the smart contract and the local IPFS to store and retrieve model weights and scores. The model weights are stored on IPFS in a serialized format, whose content IDs are then stored on the smart contract. 

The clients participating in training operate as standard Flower clients and remain unaffected by the changes made to the aggregators for \proj{}. These clients receive the aggregated global models, perform training, and submit the results. At the end of each round, the aggregated model is saved for plotting as the local model and pushed to the blockchain to be aggregated for the next round. 

\proj{} supports a range of policies for efficient aggregation by providing the global models and their respective lists of scores for the aggregator's use. Several policies are implemented to maximize performance based on the specific use case that efficiently utilizes the global models and their associated scores. A set of aggregation policies is defined to determine which models are selected for global aggregation, and a set of scoring policies is defined to specify how to pick a score from the set of scores provided by the smart contract to the blockchain. A description of these policies is available in Section~\ref{sec:policies}.

To facilitate the evaluation of \proj{} under various data distribution scenarios, we leverage the Flower Datasets library, which provides easy access to datasets from the Hugging Face dataset hub~\citep{huggingface_datasets} and enables us to obtain both independent and identically distributed (IID) and non-IID partitioned datasets using a variety of distributions. The seamless integration with the Hugging Face dataset hub allows us to experiment with a wide range of datasets, from natural language processing tasks to computer vision problems, further enhancing the versatility and applicability of \proj.

\section{Evaluation}
\label{sec:eval}

Our evaluation of the \proj framework addresses the following questions:

\begin{enumerate}
    \item\diff{  Does \proj improve the convergence of FL models in a real-world setting?}

    \item\diff{Does \proj provide flexibility of allowing different FL clusters to choose different aggregation algorithms?}
   \item\diff{Can \proj work in both NIID and IID conditions under different configurations?}
   \item \diff{How does performance of the $Sync$ and $Async$ modes of aggregation in \proj vary from one another? }
   \item\diff{Can \proj handle device heterogeneity in an edge computing scenario?}
   \item \reb{How does \proj tackle scalability?}
   \item\diff{What is the system overhead of running \proj? }
\end{enumerate}

\subsection{Experimental Setup}
\label{sec:exp_setup}

The experiments were run on two different testbeds: GPU Cluster and Edge Cluster. 
\begin{itemize}
    \item The high-performance \textit{GPU Cluster} is composed of 4 GPU nodes (i7-12700, NVIDIA RTX A2000, 64GB RAM), each that hosts the aggregator and 3 clients. A private IPFS chain was deployed on the nodes and a local Ethereum testnet was created using Anvil.
    \item  The heterogeneous and resource-constrained \textit{Edge Cluster} involves 3 CPU nodes (i7, 8GB RAM) which hosts the aggregators, Geth and IPFS. Each aggregator has a set of three homogeneous clients, consisting of Raspberry Pi 400 (4GB RAM), NVIDIA Jetson Nano (128-core Maxwell GPU, 4GB RAM), and Docker Containers (2GB RAM), respectively.
\end{itemize}




\subsubsection{Workloads} We evaluate \proj on commonly used workloads in FL literature \citep{ghosh2020efficient,fallah2020personalized,zhao2018federated,thonglek2020federated,konecny2016federated} as described in  Table~\ref{tab:exps}. The tasks include training the larger VGG16 \citep{simonyan2015deep} model with 138M parameters on the Tiny ImageNet dataset \citep{imagenet} with 200 classes (labels) on the high-performance GPU cluster to evaluate \proj{}'s ability to handle demanding workloads, and a lightweight CNN model with 62K parameters on the CIFAR-10 dataset \citep{Krizhevsky09learningmultiple} with 10 classes on the resource-constrained edge cluster. 

\begin{table}[h!]
\tiny
\caption{Configuration of workloads used in the evaluation.}
\begin{adjustbox}{width=\columnwidth}
\label{tab:exps}
\begin{tabular}{lll}
\toprule
& \thead{CIFAR-10~\citep{Krizhevsky09learningmultiple}} & \thead{Tiny ImageNet~\citep{imagenet}} \\
\midrule
\textbf{Task} & Image Classification & Image Classification \\
\textbf{Model} & CNN & VGG16~\citep{simonyan2015deep} \\
\textbf{\# of Params} & 62K & 138M \\
\textbf{Learning Rate} & 0.01 & 0.01 \\
\textbf{Rounds} & 100 & 50 \\
\textbf{Local Epochs} & 2 & 2 \\
\textbf{Batch Size} & 5 & 64 \\
\textbf{\# of Labels} & 10 & 200 \\
\textbf{Testbed} & Edge Cluster & GPU Cluster \\
\bottomrule
\end{tabular}
\end{adjustbox}
\end{table}

\subsubsection{Datasets} To model realistic heterogeneous data scenarios, we divide the labeled training data that consists of thousands of data samples among the learners using different methods - a random uniform distribution-based IID partitioning and a Dirichlet distribution-based \citep{yurochkin_dirchlet} non-IID partitioning with varying $\alpha={0.1, 0.5}$ to evaluate scenarios of uneven data availability.

\subsubsection{Hyper-parameters} We use stochastic gradient descent (SGD) as the client's local optimizer, with a learning rate of 0.01 for both image classification tasks. The CNN and VGG16 models are trained for 2 local epochs each, with batch sizes of 5 and 64, respectively. \\

This evaluation setup demonstrates the versatility of \proj{} in handling diverse models, datasets, and computing environments for FL.




\subsection{Experimental Results}

In this section, we provide experimental results to answer the questions raised before. \\

\begin{table*}[h!]
\tiny
\caption{\centering
Results of Tiny-Imagenet workload on GPU Cluster. Config defines the experiment parameters we use for the runs. Agg (1,2,3,4) rows indicate the metrics for individual FL aggregators. Time is the total time taken by each aggregator. Policy is the aggregation policy used by the respective aggregators. Local and Global metrics correspond to the locally aggregated model and \proj's globally aggregated model. }
\begin{adjustbox}{width=1.3\columnwidth}
\begin{tabular}{cc|cccccccc}
\toprule
\thead{Run} & \thead{Config} & \thead{Aggregator} & \thead{Time} & \thead{Policy} & \multicolumn{2}{c}{\thead{Accuracy}} & \multicolumn{2}{c}{\thead{Loss}} \\
 & & & & & \thead{Global} & \thead{Local} & \thead{Global} & \thead{Local} \\
\midrule
1 & HBFL Baseline & Agg 1 & 6230 & All & 36.84 & 31.07 & 2.72 & 3.22 \\ 
   &FedAvg & Agg 2 & 6230 & All & 36.84 & 31.21 & 2.72 & 3.13  \\
   &Accuracy & Agg 3  & 6230 & All & 36.84 & 31.30 & 2.72 & 3.15 \\
   & NIID $\alpha=0.5$ & Agg 4 & 6230 & All & 36.84 & 31.26 & 2.72 & 3.12 \\
\midrule
2 & \proj Async & Agg 1 & 4231 & All & 34.32 & 28.85 & 2.74 & 3.15 \\
  & FedAvg & Agg 2 & 4128 & All & 34.31 & 28.55 & 2.76 & 3.14 \\
  & Accuracy& Agg 3 & 4112 & All & 33.84 & 28.69 & 2.75 & 3.13 \\
  &NIID $\alpha=0.5$ & Agg 4 & 4084 & All & 33.83 & 28.97 & 2.75 & 3.07 \\
\midrule
3 & \proj Async & Agg 1 & 4144 & Top2 Mean & 27.20 & 22.39 & 3.29 & 3.81 \\
  & FedAvg & Agg 2 & 3472 & Top2 Mean & 25.72 & 19.61 & 3.24 & 3.54 \\
  & Accuracy & Agg 3 & 3869 & Top2 Mean & 27.13 & 22.15 & 3.18 & 3.69 \\
  & NIID $\alpha=0.1$ & Agg 4 & 4376 & Top2 Mean & 27.77 & 23.53 & 3.07 & 3.63 \\
\midrule
4 & \proj Async & Agg 1 & 4201 & Top2 Mean & 22.66 & 22.27 & 4.10 & 3.76 \\
  & FedYogi & Agg 2 F& 3520 & Top2 Mean & 24.63 & 18.82 & 3.26 & 3.65 \\
  & Accuracy & Agg 3 & 3844 & Top2 Mean & 26.51 & 21.94 & 3.26 & 3.64 \\
  & NIID $\alpha=0.1$ & Agg 4 F& 4431 & Top2 Mean & 28.74 & 22.36 & 3.17 & 4.44 \\
\midrule
5 & \proj Sync & Agg 1 & 6380 & Self & 22.11 & 22.79 & 7.21 & 7.34 \\
   & FedAvg & Agg 2& 6380 & Top2 Max & 35.03 & 31.77 & 3.46 & 3.65 \\
   & Accuracy & Agg 3 & 6380 & Top2 Mean & 35.25 & 31.53 & 3.48 & 3.92 \\
   & NIID $\alpha=0.5$ & Agg 4 & 6380 & Top3 Mean & 36.49 & 32.47 & 3.38 & 3.83 \\
\midrule
6 & \proj Sync & Agg 1 & 6258 & Self & 21.22 & 21.31 & 7.02 & 7.55 \\
  & FedAvg & Agg 2 & 6258 & Top2 Max & 33.02 & 31.93 & 3.18 & 3.43 \\
  & Accuracy& Agg 3 & 6258 & Top2 Mean & 32.49 & 31.59 & 3.55 & 3.36 \\
  & IID & Agg 4 & 6258 & Top3 Mean & 35.10 & 32.32 & 3.10 & 3.48 \\
\midrule
7 & \proj Sync & Agg 1 & 6264 & All & 34.71 & 30.59 & 2.89 & 2.98 \\
  & FedAvg& Agg 2 & 6264 & Top3 Mean & 31.75 & 27.88 & 3.45 & 3.96 \\
  & MultiKRUM& Agg 3 & 6264 & Top2 Mean & 33.10 & 29.07 & 3.07 & 3.66 \\
  & NIID $\alpha=0.5$ & Agg 4 & 6264 & Top1 Mean & 27.60 & 29.11 & 3.89 & 3.05 \\
\midrule
8 & \proj Sync & Agg 1 & 6391 & All & 36.99 & 33.89 & 2.84 & 3.03 \\
  & FedAvg & Agg 2 & 6391 & All & 37.10 & 34.90 & 2.83 & 3.03 \\
  & Accuracy & Agg 3 & 6391 & All & 36.73 & 35.11 & 2.85 & 3.03 \\
  & IID & Agg 4 & 6391 & All & 37.14 & 34.81 & 2.84 & 3.04 \\
\midrule
9 & \proj Async & Agg 1 & 4258 & All & 38.83 & 36.25 & 2.77 & 2.99 \\
   & FedAvg & Agg 2 & 4053 & All & 37.59 & 35.91 & 2.84 & 2.88 \\
   & Accuracy & Agg 3 & 4157 & All & 37.99 & 35.67 & 2.83 & 3.04 \\
   & IID & Agg 4 & 4068 & All & 36.04 & 35.38 & 2.85 & 3.11 \\\bottomrule
\label{tab:imagenet_eval}

\end{tabular}
\end{adjustbox}
\end{table*}

\textbf{Baseline: } To set a comparative benchmark for \proj, we implement HBFL~\citep{sarhan2022hbfl}, a blockchain-based hierarchical FL framework. This framework replicates the optimal scenario of centralized multilevel FL outlined in Section~\ref{sec:central_FL} \reb{and serves as an oracle baseline}, utilizing a NIID-partitioned dataset on CIFAR-10. In Table~\ref{tab:collab-nocollab}, it is evident that the model's global accuracy surpasses the local accuracy of any aggregator across 50 FL rounds. This trend is consistent in the NIID-partitioned dataset on ImageNet, as illustrated in Table \ref{tab:imagenet_eval} Run 1.
   

\subsubsection{\diff{ Does \proj improve the convergence of FL models in a real-world setting?} }

\diff{  Table \ref{tab:imagenet_eval} Run 2 } shows the accuracy and loss of using Async \proj using the same experimental setup as baseline. \textit{Pick All} policy is used to select the set of models. It is seen that \proj achieves comparable accuracy (35\%) to the baseline (Table \ref{tab:imagenet_eval} Run 1 ) while demonstrating significantly faster runtime (4000 secs vs 6000 secs) over 50 FL rounds. The asynchronous nature of \proj{} contributes to its speed, providing a notable improvement over the traditional synchronous approach.

\subsubsection{ \diff{ Does \proj provide flexibility of allowing different FL clusters to choose different aggregation algorithms? } }

\proj{} demonstrates its flexibility by allowing different FL clusters to seamlessly integrate and leverage their preferred aggregation algorithms. This flexibility is demonstrated through a comparative analysis of the FedAvg~\citep{mcmahan2017communication} and FedYogi~\citep{fedyogi} algorithms in an asynchronous NIID ($\alpha=0.1$) setting. In \diff{Table \ref{tab:imagenet_eval} Run 3}, all aggregators adopt the FedAvg algorithm, whereas in \diff{Table \ref{tab:imagenet_eval} Run 4}, Aggregators 1 and 3 employ FedAvg while Aggregators 2 and 4 utilize FedYogi. \reb{
Moreover, \diff{Table \ref{tab:imagenet_eval} Run 5} highlights aggregators adopting distinct aggregation policies tailored to their specific needs. These experiments illustrate how different aggregation strategies can coexist within the same FL framework without disrupting existing FL pipelines. By decoupling aggregation logic from the framework’s core functionalities, \proj{} facilitates a plug-and-play approach, allowing clusters to optimize performance based on their unique data distributions, computational resources, and constraints.
}
\subsubsection{\diff { Can \proj work in both NIID and IID conditions under different configurations?} }

To illustrate the performance of \proj{} across various aggregation strategies and dataset partitioning, we consider a setup with four clusters, each of them with a unique policy as follows: \textit{Pick Self}, \textit{Pick Top 2 by Max Score}, \textit{Pick Top 2 by Mean Score}, \textit{Pick Top 3 by Mean Score}. Notably, \textit{Pick Self} policy indicates the aggregator choosing not to collaborate. \diff{ Table \ref{tab:imagenet_eval} Run 5}
shows the accuracy and loss of using Sync \proj using the above experimental setup on a NIID partitioned dataset. It is seen that \proj achieves comparable accuracy (35\%) to the baseline (Table \ref{tab:imagenet_eval} Run 1 ) cross collaborative policies while demonstrating similar runtime over 50 FL rounds. Moreover, aggregator 1 choosing not to collaborate has a much lower accuracy (22\%), indicating the importance of collaborative FL and \proj.
Similarly, \diff{ Table \ref{tab:imagenet_eval} Run 6 }
shows that Sync \proj can reach an accuracy (33-35\%) on par with a single-level FL setup (37\% in 75 rounds) with 12 clients, showcasing its competence under IID conditions irrespective of the aggregation policy employed. These results conclusively establish the versatility of \proj{}, validating its ability to adapt seamlessly to diverse data distribution scenarios without compromising performance.
To demonstrate that \proj{} supports other scoring algorithms, we have also run the sync \proj on a NIID dataset scored by the $MultiKRUM$ algorithm. 
\diff{ Table \ref{tab:imagenet_eval} Run 7}
shows similar performance to 
\diff{ Table \ref{tab:imagenet_eval} Run 6}.

\begin{table*}[h!]
\tiny
\caption{\centering Results of CIFAR-10 workload on Edge Cluster. Config defines the experiment parameters we use for the runs. Agg (1,2,3,4) rows indicate the metrics for individual FL aggregators. Time is the total time taken by each aggregator. Policy is the aggregation policy used by the respective aggregators. Local and Global metrics correspond to the locally aggregated model and \proj's globally aggregated model.}
\begin{adjustbox}{width=1.3\columnwidth}
\begin{tabular}{cc|cccccccc}
\toprule
\thead{Run} & \thead{Config}  & \thead{Aggregator} &  \thead{Time} & \thead{Policy} & \multicolumn{2}{c}{\thead{Accuracy}} & \multicolumn{2}{c}{\thead{Loss}} \\
 & & & & & \thead{Global} & \thead{Local} & \thead{Global} & \thead{Local} \\
\midrule
C1 & \proj Sync  & Agg 1 & 4460 & Top2 Mean & 59.83 & 59.16 & 1.13 & 1.15 \\
  &  FedAvg  & Agg 2 & 4460 & Top2 Mean & 59.83 & 59.77 & 1.13 & 1.14 \\
  &  Accuracy  & Agg 3 & 4460 & Top2 Mean & 59.83 & 59.30 & 1.13 & 1.13 \\
  & IID  & & & & & & & \\
\midrule
C2 & \proj Sync   & Agg 1 & 4420 & Top2 Mean & 51.33 & 30.54 & 1.77 & 4.92 \\
  &  FedAvg  & Agg 2 & 4420 & Top2 Mean & 51.33 & 34.54 & 1.77 & 4.73 \\
  &  Accuracy   & Agg 3 & 4420 & Top2 Mean & 51.33 & 31.43 & 1.77 & 4.18 \\
  & NIID $\alpha=0.5$  & & & & & & & \\
\midrule
C3 & \proj Async  & Agg 1 & 2455 & Top2 Mean & 44.58 & 26.85 & 2.17 & 4.63 \\
  &  FedAvg  & Agg 2 & 2093 & Top2 Mean & 43.08 & 33.09 & 2.19 & 5.22 \\
  &  Accuracy  & Agg 3 & 3209 & Top2 Mean & 44.72 & 29.87 & 2.16 & 5.35 \\
  & NIID $\alpha=0.5$ &  & & & & & & \\
\bottomrule
\label{tab:cifar_eval}

\end{tabular}
\end{adjustbox}
\end{table*}

\subsubsection{\diff{ How does performance of the $Sync$ and $Async$ modes of aggregation in \proj vary from one another?} }

The choice between synchronous (Sync) and asynchronous (Async) orchestration modes in \proj{} has a significant impact on the convergence speed of the FL process. To evaluate this, we use the IID partitioned dataset on the GPU cluster with all 4 aggregators selecting the pick-all policy. \diff{ Table \ref{tab:imagenet_eval} Run 8 and Run 9 }
show the results of Async and Sync modes respectively. The Async mode achieves substantially faster convergence in terms of accuracy compared to the Sync mode (4000 seconds versus 6000 seconds). This superior performance can be attributed to the Async mode's ability to minimize idle time within the cluster, ensuring that available computational resources are consistently utilized for training and model updates.
In the Sync mode, the entire cluster must wait for a fixed amount of time, waiting for the slowest cluster, before proceeding to the next round of training, leading to potentially underutilized resources. Conversely, the Async mode decouples the progress of individual clusters, allowing faster devices to contribute more frequent model updates without being hindered by slower counterparts, thus maximizing resource utilization, accelerating convergence, and improving overall efficiency. The results are similar for NIID dataset with other aggregation policies, thereby proving the efficacy of \proj for handling stragglers.



\subsubsection{ \diff{ Can \proj handle device heterogeneity in an edge computing scenario?} }

To evaluate the practical applicability of \proj{} in resource-constrained environments, we deploy it on an edge cluster comprising of heterogeneous hardware configurations with varying computational capabilities. Employing a lightweight workload involving a Convolutional Neural Network (CNN) and the CIFAR-10 dataset, we examine \proj{}'s performance under both Sync and Async modes in a NIID scenario, as illustrated in \diff{  Table \ref{tab:cifar_eval} Run C2 and Run C3, }
respectively. Sync \proj achieves an accuracy (51\%) slightly beating the centralized FL (50\%), as depicted in  
\diff{ Table \ref{tab:collab-nocollab} }
Async \proj reaches a lower accuracy (44\%) due to the limited availability of models in async mode but at a significantly lower runtime (2500 secs vs. 4400 secs). \proj also attained an acceptable level of accuracy in an IID scenario, as shown in \diff{ Table \ref{tab:cifar_eval} Run C1. }
These findings highlight \proj{}'s suitability and robustness for real-world deployments, where heterogeneous resource-constrained hardware and varying data distributions are common.

\subsubsection{\reb{How does \proj{} tackle scalability?}}
\reb{
Scalability is essential for real-world FL deployments. In our experiments with 60 clients split between 3 aggregators, \proj{} maintained stable performance trends achieving an accuracy (\~30\%) comparable to the baseline for a similar configuration at the end of 100 rounds, aligning with the previous results. Compared to a two-tier hierarchical FL system, where edge clients often lack the computational resources to run blockchain and IPFS nodes, \proj{} abstracts these complexities at the cluster level. This design enables decentralized orchestration without burdening individual edge clients while allowing larger federations to operate efficiently. These results highlight \proj{}'s ability to scale across different client sizes while preserving flexibility and performance.
}

\subsubsection{What is the system overhead of running \proj?\nopunct} Our evaluations show that the resource utilization incurred while running the Geth chain and IPFS is minuscule compared to that of the FL workload. In our experimental setup, Geth uses 0.2\% CPU and 6 MB of memory while IPFS utilizes 3.5\% CPU and 19 MB of memory. 
We also report the overall usage of system resources by various actors while performing FL tasks in \proj in \diff{ Table~\ref{tab:metrics}}. 
\reb{Notably, this overhead remains constant even when scaling up to 60 clients, demonstrating that \proj{} does not introduce additional computational burden as the system scales}. The system utilization by the FL tasks shows that running \proj incurs negligible system overhead compared to the FL model. 

\begin{table}[h!]
\scriptsize
\caption{Systems metrics of Aggregators and Clients in \proj{}.}
    \begin{tabular}{llrr}
\toprule
Process & Type & \multicolumn{2}{c}{\proj{}} \\
        &      & Mean & Std/Dev \\
\midrule
scorer & cpu \% & 11.425 & 28.598 \\
       & mem (MB) & 1038.219 & 66.395 \\
\midrule
agg    & cpu \% & 4.056 & 15.010 \\
       & mem (MB) & 11387.396 & 4487.669 \\
\midrule
client & cpu \% & 61.400 & 48.468 \\
       & mem (MB) & 1810.481 & 71.348 \\
\bottomrule
\end{tabular}    
    \label{tab:metrics}
\end{table}

\section{Limitations and Future Work}
\label{sec:discussion}

In this section, we answer a few questions and discuss the limitations of \proj.\newline

\textbf{Q1. Does \proj{} allow flexibility in training different FL models in different FL clusters?}\\
Enabling multi-model FL, where organizations can collaborate while operating on different models, would allow for greater flexibility and heterogeneity in the FL ecosystem, catering to diverse use cases and organizational requirements. \proj in its present state only supports single-model FL but offers the flexibility of selecting multiple aggregation algorithms. By supporting multi-model FL in the future \diff{ through shared layers ~\citep{thapa2022splitfed} ~\citep{Bhuyan_2022} or knowledge distillation ~\citep{muhammad2021robust} ~\citep{yu2022resource} ~\citep{li2022feddkdfederatedlearningdecentralized}, \reb{as well as leveraging multi-modal models and multi-modal fusion techniques~\citep{fed-multimodal}~\citep{fedmsplit} where different aggregators handle different modalities and merge them effectively}, } \proj{} will be able to encourage cross-organizational collaboration, where clusters with varying model architectures can contribute to a shared learning objective while preserving their models and infrastructures.\newline

\textbf{Q2. Can \proj{} protect against bad actors?}

\begin{figure}[h!]
\centering
    \begin{subfigure}{0.235\textwidth}
        \includegraphics[width=\textwidth]{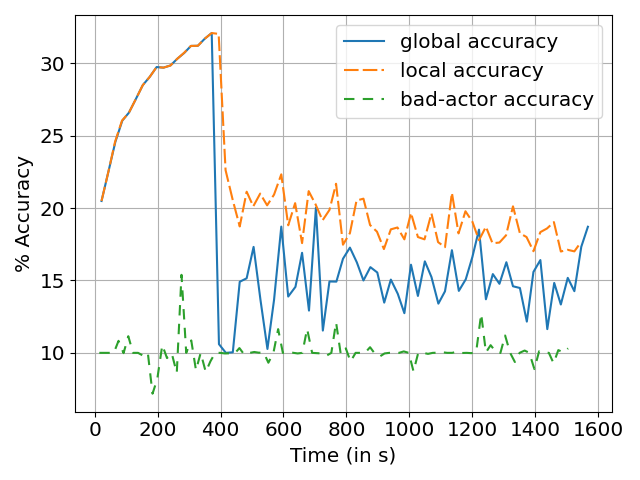}
        \subcaption{Naive Policy.}
    \end{subfigure}
    \begin{subfigure}{0.235\textwidth}
        \includegraphics[width=\textwidth]{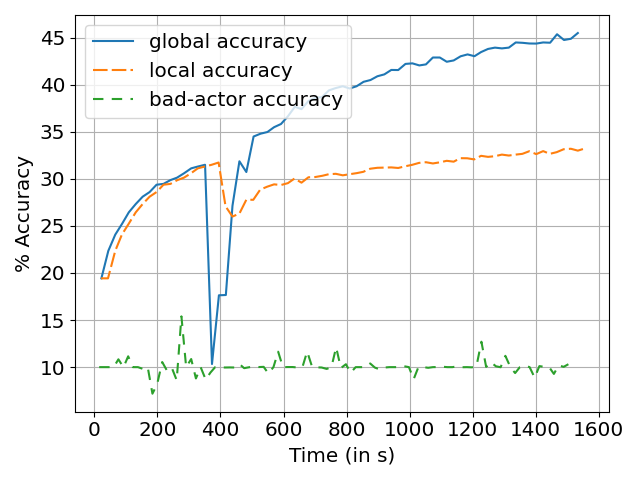}
        \subcaption{Smart Policy.}
    \end{subfigure}
    \caption{Policies to prevent Byzantine attacks.}
     \label{dis:byzantine}
\end{figure}

Protection against Byzantine attacks~\citep{shi2022challenges} is crucial in FL systems, as malicious actors can degrade global model performance. \diff{Byzantine attacks occur when malicious participants deliberately introduce incorrect or adversarial updates.} A naive policy that includes untrusted models can hinder accuracy due to adversarial noise or poisoning attacks. In contrast, a smart policy filters out suspicious models, mitigating these threats. Figure~\ref{dis:byzantine} illustrates an adversarial scenario where a bad actor submits malicious models alongside two honest aggregators. In the first 30 rounds (400 secs), each aggregator picks its own model for training, introducing reduced randomness during later aggregation. This causes a temporary accuracy dip before recovery.

In Figure~\ref{dis:byzantine}(a), the naive policy selects the top 3 global models for aggregation without considering reliability, allowing malicious contributions. In contrast, Figure~\ref{dis:byzantine}(b) shows a smart policy that aggregates only above-average models, effectively excluding bad actors. As a result, \proj{} demonstrates resilience, with the smart policy achieving significantly better accuracy. \reb{However, the overall accuracy is lower compared to previous results since training effectively occurs on only two organizations' data, as the third is malicious.}
\diff{Exploration of more sophisticated adversarial attacks, where malicious participants introduce subtle but harmful changes to model updates or adaptive attacks that exploit specific weaknesses in the aggregation strategy or the training pipeline, poses significant challenges similar to traditional federated learning. However, we plan to address these threats in future research to enhance \proj{}'s resilience. Their integration will be a valuable direction for future work.}\newline

\textbf{Q3. Does \proj{} guarantee data and model privacy?}\\ \proj{} currently provides the same data and model privacy guarantees as traditional FL. The training data remains distributed across client devices and prevents the server from directly accessing individual client data, ensuring data privacy. \proj{} does not currently implement additional privacy-enhancing techniques like Differential Privacy (DP)~\citep{10.1007/11681878_14} - introduces controlled noise to the model updates, Homomorphic Encryption (HE)~\citep{yi2014homomorphic} - allows computations to be performed directly on encrypted data, or Secure Multi-Party Computation (SMPC)~\citep{10.1145/237814.238015} - enables multiple parties to collaboratively compute a function, which could further strengthen the privacy guarantees of the system. Future work for \proj{} will incorporate techniques like DP, HE, and SMPC into \proj to enhance the system's privacy guarantees beyond traditional FL.

\section{Conclusion}
\label{sec:conclusion}

In this work, we have developed \proj that enables collaborative learning among diverse organizations in a federated learning (FL) setup. \proj ensures trust among multiple aggregators via a decentralized orchestration while achieving similar accuracy convergence to multilevel FL which uses centralized orchestration. We also devised strategies to handle failures and stragglers by operating in two modes - Sync and Async. Our evaluation on a real testbed shows the efficacy of \proj with minimal resource utilization. \proj also offers flexibility to different FL aggregators to select their custom algorithms and aggregation policies. As part of future work, we will extend \proj to incorporate policies for handling malicious aggregators, integrate privacy-preserving techniques, and enable multi-model FL training.

The repository with the code, experimental results, and installation instructions is provided at \url{https://github.com/DaSH-Lab-CSIS/UnifyFL}.

\begin{acks}
This work was supported in part by BITS Pilani CDRF grant C1/23/173, BioCyTiH grant BBF/BITS(G)/FY2022-23/BCPS-123/24-25/R1, and SERB SRG grant SRG/2023/002445. We also thank the contributions of Pinki Yadav, a postdoctoral research associate in the Data, Systems and HPC Lab (DaSHLab) in BITS Pilani, KK Birla Goa Campus, India, for her inputs during the project ideation phase.
\end{acks}

\bibliographystyle{ACM-Reference-Format}
\bibliography{refs}

\end{document}